\documentclass[nohyper,12pt,letterpaper]{JHEP}
\usepackage{epsfig}
\newfont{\frak}{eufm10 scaled 1200}

\newfont{\Bbb}{msbm10 scaled 1200}     %instead of eusb10
\newcommand{\mathbb}[1]{\mbox{\Bbb #1}}
\DeclareSymbolFont{AMSa}{U}{msa}{m}{n}
\DeclareSymbolFont{AMSb}{U}{msb}{m}{n}
\let\Box\relax
\DeclareMathSymbol{\Box}{\mathord}{AMSa}{"03}

\title{A Dark Matter Candidate With New Strong Interactions}
\author{T. Banks\\
    Department of Physics and Astronomy - NHETC\\
    Rutgers University\\
    Piscataway, NJ 08540\\
    and\\
    Department of Physics, SCIPP\\
    University of California, Santa Cruz, CA 95064\\
E-mail: \email{banks@scipp.ucsc.edu}}
\author{J.D. Mason and D. O'Neil\\
    Department of Physics, SCIPP\\
    University of California, Santa Cruz, CA 95064\\
E-mail: \email{doneil@physics.ucsc.edu, jdmason@physics.ucsc.edu}}

\abstract{We study the possibility that dark matter is a baryon of
a new strongly interacting gauge theory, which was introduced in
the low energy theory of Cosmological SUSY Breaking (CSB).  This
particle can fit the observed dark matter density if an
appropriate cosmological asymmetry is generated.  The same
mechanism can also explain the dark/baryonic matter ratio in the
universe.  The mass of the dark matter particle is in the multiple
TeV range, and could be as high as 20 TeV.}

\keywords{dark matter, supersymmetry}

\preprint{\hepph{0506015}\\SCIPP-11/05}

\begin{document}

%%%%%%%%%%%%%%%%%%%%%%%%%%%%%%%%%%%%%%%%%%%%%%%%%%%%%%%%%%%%%%%%%%%%%%%%%%%%

%          Table of contents automatic !!!                                 %

%%%%%%%%%%%%%%%%%%%%%%%%%%%%%%%%%%%%%%%%%%%%%%%%%%%%%%%%%%%%%%%%%%%%%%%%%%%%

\section{\bf Introduction}

The most common particle physics models for dark matter involve
weakly interacting particles.   They can be broadly classified as
WIMPS or axions, with the theoretician's favorite WIMP being a
neutralino of the Supersymmetric Standard Model (SSM).   Within
string theory, the physics of both of these candidates is closely
connected to SUSY breaking, because string theory axions generally
arise from moduli fields, whose mass is related to a
superpotential on moduli space.

One of the authors has recently introduced a new model for SUSY
breaking, which has no candidate for either WIMP or axion dark
matter\cite{susycosmopheno}.   The model is based on the principle
of Cosmological SUSY Breaking (CSB):

\begin{itemize}

\item The (positive) cosmological constant (c.c.) is a discrete tunable
parameter, governing the number of states in the Hilbert space of
quantum gravity in de Sitter (dS) space.

\item As the c.c. vanishes, SUSY is restored, with the relation
$m_{3/2} \sim \Lambda^{1/4} $ between the gravitino mass and the
c.c.  A discrete $Z_n$ $R$ symmetry is restored in the same limit,
explaining, in low energy terms, the vanishing of the c.c. in the
SUSic limiting theory.  The limiting theory must have a compact
moduli space, in order to guarantee that the dS state of the low
energy effective field theory is stable.

\item SUSY breaking is spontaneous in the low energy effective
theory, but is induced by $R$ breaking terms in the Lagrangian
which have no low energy explanation.  The coefficients in these
terms are tuned to guarantee the CSB scaling relation between
$m_{3/2}$ and $\Lambda$.

\end{itemize}

As a consequence of the first requirement, the low energy
effective field theory of CSB must contain a goldstino field: a
linear supermultiplet which is massless in the SUSic, R symmetric
limit. In \cite{susycosmopheno} this was taken to be a chiral
superfield $G$, with $R$ charge $0$.   If there are no fields of
$R$ charge $2\ mod\ n$ in the low energy theory, then $G$ is
naturally massless. $R$ charges were assigned to standard model
fields in a way that insured the absence of all baryon and lepton
number violating dimension $4$ and $5$ operators, apart from the
term $n^{ij} H_u^2 L_i L_j$ (which gives rise to neutrino masses).
The generation of this term, and of the texture of Yukawa
couplings is imagined to have to do with physics at the
unification scale.   There is also an ordinary discrete symmetry
${\cal F}$, under which $G$ transforms. ${\cal F}$ allows the
coupling $g_{\mu} G H_u H_d $ but forbids the conventional $\mu$
term.  $G^a$ is the lowest order ${\cal F}$ invariant monomial in
$G$.

High energy physics supplies us with a term $M_P^2 \Lambda^{1/4}
f(G/M_P)$ which violates $R$ and implements CSB.   The
dimensionless coefficients in the function $f$ are tuned to
guarantee that the c.c. is indeed $\Lambda$.  For phenomenological
reasons, one must also add terms
$$\int d^4 \theta M_1^2 K(g,h_u, h_d, q,\bar{u},\bar{d},l,
\bar{e} ),$$ and,
$$\int d^2 \theta Z_A (g^a ) W_A^2 + h.c. .$$
We have used an unconventional notation where a lower case label
$s$ for a chiral superfield $S$ stands for $S/M_1$.   The Kahler
potential depends, of course, both on chiral fields and their
conjugates.   The functions $K$ and $Z_A$ are imagined to emerge
from integrating out degrees of freedom at a scale $M_1 \ll M_U
\ll M_P$, whose value is determined by RG flow in the limiting
$\Lambda = 0$, theory.   They can be chosen to satisfy all
phenomenological requirements if $M_1 \sim 1$ TeV.   It is easy to
invent strongly coupled theories ${\cal G}$ which could give rise
to all the required properties save one.   There is no known
example of a theory which preserves the $R$ symmetry, and leaves
exactly one effective chiral superfield which could play the role
of $G$. We will leave this problem to future work and concentrate
on the problem of dark matter.

If the coupling functions $Z_A$ were forced to be logarithms by an
accidental $U(1)$ with standard model anomalies, then the real
part of $G$ could be a QCD axion.   However, it would have a range
of axion couplings ruled out by beam dump experiments.
Consequently the model has no axion candidates.   The basic setup
of CSB contradicts the idea of SUSY neutralino dark matter.  The
gravitino is the LSP in the CSB scenario, and its longitudinal
components are relatively strongly coupled, so the NLSP is not
cosmologically stable.

The only plausible dark matter candidate in this scenario is what
we will call a ${\cal G}$ baryon.   That is, we assume the
strongly interacting ${\cal G}$ sector has an accidental symmetry,
which renders the lightest particle carrying some accidental
$U(1)$ quantum number, cosmologically stable. In this paper, we
will explore the idea that the dark matter is in fact a baryon of
a strongly interacting sector with an RG scale of order $M_1$. We
will see that under a variety of assumptions about the production
of this particle, this hypothesis is consistent with conventional
cosmology. It has the added virtue of correlating the coincidence
between the dynamical scale $M_1$ and the CSB scale
$\sqrt{(\Lambda^{1/4} M_P)}$ to the existence of galaxies.   That
is to say, we imagine that the limiting model calculates the value
of the scale $M_1$ and the other parameters of ${\it e.g.}$ the
inflaton field, in such a way that the density of ${\cal G}$
baryons coincides with what we know about dark matter density from
observations.   Now consider the model of CSB, with various values
of $\Lambda$.   The only values which will produce a model with
galaxies will be those which satisfy Weinberg's bound.  At least
within a few orders of magnitude, this matches the scale of CSB to
$M_1$ and the dark energy density to the dark matter density
(cosmic coincidence).

We will also see that there is a variety of thermal histories for
the universe in which ${\cal G}$ baryons can be dark matter only
if there is a CP violating ${\cal G}$ baryon number asymmetry. We
might imagine a model in which ${\cal G}$ and ordinary baryon
asymmetries were produced by the same mechanism, perhaps
explaining the dark/baryonic matter ratio of the
universe\cite{witten}\cite{kaplan}.

The Hess telescopes\cite{hess} have seen a photon signal from the
center of the galaxy, which might be consistent with a dark matter
candidate of mass $15-18$ TeV, if dark matter in the galaxy
follows the profile predicted by \cite{joel}. It is very hard to
find a neutralino model which can produce such a large mass,
basically because weak annihilation cross sections decrease with
mass.   On the other hand, strongly interacting particles have
mass independent annihilation cross sections and can easily fit
this data.

In the next section, we estimate various cross sections for baryon
like objects, using large $N$ QCD as a paradigm. The ${\cal G}$
theory must differ from QCD since it preserves chiral symmetry and
is supersymmetric.   Nonetheless, we hope that these estimates
give us a rough guide to the scales involved.  We then go on to
estimate the mass, cross section and primordial asymmetry for
which a ${\cal G}$ baryon could be dark matter.  We consider two
scenarios: a standard thermal relic abundance calculation, and a
particular non-thermal production scheme.   We find that for
reasonable values of parameters, the model can fit the data, and
perhaps reproduce the Hess signal.   To answer the latter question
in more detail, one must perform a detailed estimate of the photon
spectrum one would get from annihilation processes involving a
strongly interacting dark matter candidate.  We are not sure that
the model used by the Hess collaboration in order to extract the
parameters of a hypothetical dark matter particle from their
signal, takes into account the physics of a strongly interacting
particle.

We should emphasize that despite our original motivation, our
calculations would be applicable to any dark matter candidate with
new strong interactions of the right scale. In particular, we note
that our model for dark matter is similar to the hypothesis that
dark matter is a techni-baryon\cite{Nussinov}\cite{Bagnasco}.

\section{Annihilation Cross Sections for Dark Matter With New Strong
Interactions}

The nucleon anti-nucleon annihilation cross section is usually
written in units of the pion Compton wavelength, because this is
the range of nuclear forces.   In fact, this parametrization is
singular in the chiral limit, when the pion becomes a Goldstone
boson. It is not correct that the cross section blows up in this
limit.

A better estimate is obtained by thinking about chiral soliton
models of the nucleon\cite{skyrmeanw}.   In such models the
nucleon is realized as a classical solution of a large $N$
effective action. The effective Planck constant of this action is
of order $N$, and the scale over which solutions vary is the QCD
scale. Although these models use the spontaneously broken chiral
symmetry of QCD in an essential way, they give the same order of
magnitude results one would expect from general large $N$
considerations. We expect the size of a general large $N$ soliton
to be given by such an $N$ independent scale, and large $N$
soliton masses will be of order $N$.

The soliton-anti-soliton annihilation cross section will be given
by its classical size $\sigma \sim  \Lambda_{\cal G}^{-2}$ and
will be more or less energy independent in the regime of interest,
because the cosmological velocities of these heavy particles will
be low.    Note that this is {\it not} s-wave annihilation. The
typical orbital angular momentum involved in these collisions is
of order ${\sqrt{m_{\cal G} T}\over \Lambda_{\cal G}}$, where $T$ is the
temperature at which the annihilation takes place.  Note also that
the thermally averaged cross section $<\sigma v>$, which appears
in cosmological Boltzmann equations, will be $O(T/m_{G})^{1/2}$.
We believe that this is the correct scaling even for ordinary
baryons, and that conventional calculations of the relic baryon
density in a baryon symmetric universe are not quite correct.
However, this does not change the qualitative conclusion of those
calculations, namely that we need a baryon asymmetry to account
for the observed baryon number density of the universe.

We note that the reason that we are interested in large $N$
counting is the combination of the Hess data, and the constraints
on $\Lambda_{\cal G}$ from supersymmetric phenomenology.   The
latter prefers a scale $\Lambda_{\cal G} \sim 1$ TeV, in order to
accommodate the bounds on charged superpartner masses, while the
former indicates a mass around $15- 18$ TeV for the dark matter
particle.  In a large $N$ model, the baryon mass would be $N
\alpha \Lambda_{\cal G}$ with $\alpha$ a number of order $1$
($\alpha \approx 2$ in QCD).   Thus, we would want $N \alpha \sim
15 - 18$. These are not unreasonable values.  For example, the
best of the inadequate models for the ${\cal G}$ theory, studied
in \cite{susycosmopheno} was an $SU(4)$ SUSY gauge theory.   For
$N =4$, we require $\alpha \sim 4$, about twice the value in QCD.

We emphasize however that we do not know the details of the model
which the Hess collaboration used in quoting $15 - 18$ TeV for
their best fit to the dark matter candidate. In particular, for
weakly coupled neutral dark matter, the direct photon annihilation
signal is suppressed by a power of $(\alpha /\pi)^2
$ relative to
photons produced from decays of particles with direct coupling to
the dark matter.   There is no such suppression for strongly
interacting neutral composites of charged particles. For example
the large $N$ nucleon magnetic moment is order $e (=
\sqrt{4\pi\alpha_{em}}) N$ in $\Lambda_{QCD}$ units.  Hess has not
yet seen the characteristic turnover in their photon signal, which
would be expected from dark matter annihilation, and the question
of astrophysical explanations for the signal from the galactic
center is still controversial.  It is perhaps premature to try to
fit their spectrum.

However, it is clear that in order to really confront an eventual
dark matter signal from Hess data, we need a much better estimate
of the photon spectrum produced by a ${\cal G}$ baryon. In addition, since we find that for most values of the reheat temperature of
the universe, we must invoke a ${\cal G}$ baryon asymmetry to
account for the observed dark matter density, the annihilation signal will
be proportional to the small density of anti-${\cal G}$ baryons.  We have
not yet done the calculations to determine the range of parameters for which
we would expect a significant annihilation signal from
the center of the galaxy. In the rest of this paper, we will choose an annihilation cross section
of order $\Lambda_{\cal G}^{-2}$ and parametrize our results in
terms of the ${\cal G}$ baryon mass $m_{\cal G} > \Lambda_{\cal
G}$, $\Lambda_{\cal G}$, and an asymmetry.

Our description of the ${\cal G}$ baryon will utilize the following
characteristics of a soliton model: energy independent
annihilation cross section much larger than the scale of its
Compton wavelength, and thermal production at energies well below
its mass.  The latter is a well known
characteristic\cite{dinefischlerpolchinski} of solitons in weakly
coupled field theory.  Finally, we will parametrize the ${\cal G}$ baryon mass as $N \alpha \Lambda_{\cal G}$, with $\Lambda_{\cal G} \sim 1$ TeV, in order to suggest the large $N$ scaling of soliton
masses in strongly coupled gauge theories with large gauge groups.
\section{\bf The Relic Abundance of ${\cal G}$ baryons}

We will denote by $\Omega_G$ the fraction of the observed density
of the universe in ${\cal G}$ baryons plus anti-baryons.  To match
the observed dark matter abundance, we require $\Omega_G \equiv
{\rho_G \over \rho_{cr}} = .24$, using the data from WMAP which
specifies $\Omega_m = .29 \pm .07$ and $\Omega_b = .047 \pm
.006$.\cite{spergel} If $n_G$ is the number of ${\cal G}$ baryons
per comoving volume, then this can be written $\Omega ={ {n_G
m_{\cal G} }\over {{3 H_0^2/ 8\pi G}}} ={ {n_G m_{\cal G} } \over
{1.054 h^2 \cdot 10^4{ eV \over cm^3}}}$.

Writing today's value of the ${\cal G}$ baryon abundance (the
ratio of the number of ${\cal G}$ baryons per comoving volume and
the entropy) $Y_0 \equiv{ n_G \over s_0}$, this condition becomes
\[\Omega = .24 ={{s_0 Y_0 m_{\cal G} } \over {1.054 h^2 \cdot 10^4 {eV \over
{cm^3}}}}\] Thus we require $Y_0 ={.44  eV \over m_{\cal G} }$.
We will write $m_{\cal G}  = N \alpha$ TeV, treating $1$ TeV as
the analog of the QCD scale for the ${\cal G}$ gauge theory, and
applying a large $N$ scaling rule for baryon masses. In QCD $N =
3$ and $\alpha \sim 2$.  Our point is that the analog of a baryon
mass could be quite a bit higher than $1$ TeV.  For example $N =
6$ and $\alpha \sim 3$ would give us an $18$ TeV dark matter
candidate, as would be required by the interpretation of Hess data
in terms of dark matter annihilation. With this parametrization,
the required value of the abundance is
  $Y_0 \approx {4\cdot 10^{-13} \over {N \alpha}}$.

The relic abundance of ${\cal G}$ baryons depends on some
assumptions about the evolution of the universe at the TeV scale
and above.  We assume that there was a reheating process which
gives rise to a radiation dominated universe at some temperature
$T_{RH}$.  This might be due to primordial inflaton decay, or the
later decay of some other massive particle which dominates the
energy density before it decays. We call the width of the particle
$\Gamma_X$ . If $T_{RH}
> 1$ TeV, the ${\cal G}$ gauge theory is thermalized by X-decay
and the post-decay distribution of ${\cal G}$ baryons is given by
the thermal ensemble.  Note that this is true even when $m_{\cal
G} \gg T_{RH}$.   In this regime of parameters, the ${\cal G}$
baryon is a thermal relic, and we find that, in the absence of an
asymmetry, the relic abundance is too small to explain the
observed dark matter density.

For $T_{RH} < 1 $ TeV, ${\cal G}$ baryons are produced
non-thermally and we must be a bit more specific about the
dynamics.   For a weakly coupled $X$ particle, $m_X \gg T_{RH}$
and we can still have $X$ decays into ${\cal G}$ baryons.  Suppose
first that $m_X \gg m_{\cal G}$ so that we can treat the ${\cal
G}$ baryons as just another massless species.  If we assume the
couplings to ${\cal G}$ baryons are not suppressed relative to
standard model particles we get a branching ratio of order
$10^{-2}$ into ${\cal G}$ baryons.  The decay will be reasonably
rapid, so we neglect annihilation processes during the decay
period and obtain an initial abundance of
$$Y_0 \sim 10^{-2} {T_{RH}\over m_{\cal G}}.$$
If the ${\cal G}$ baryon were massless, this ratio would just be
the branching ratio $10^{-2}$.  The additional suppression is our
estimate of the number of ${\cal G}$ baryons per photon that
result from the thermalization process.

Throughout the interesting range of parameters, the X particle
life-time is short enough to neglect annihilation in the
calculation above. Now we can evolve the resulting ${\cal G}$
baryon densities according a Boltzmann equation driven only by the
annihilation of $g$ and $\bar{g}$. As in the discussion of
solitonic dark matter abundance in Griest and Kamionkowski
\cite{Griest:1989wd}, the thermally averaged annihilation cross
sections have temperature dependence given by : $<\sigma |v|> =
\sigma_0(\frac{T_{RH}}{m_{\cal G} })^{1/2}$. Also,we will assume
that $T_{RH} < m_{\cal G}$. In this case there
can be no process in the Boltzmann equation that creates ${\cal
G}$ baryons because it is not energetically favorable. The
Boltzmann equation for the evolution of ${\cal G}$ baryons is:

\[\dot{n}_g + 3Hn_g = -<\sigma |v|>n_g^2 \]

Letting $Y \equiv n_g/s$ and $x \equiv m_{\cal G} /T$ we get

\[ {dY \over dx}  = -{x^{1/2}\sigma_0 sm_{pl} \over 1.67g_\ast^{1/2}m_{\cal G} ^2}Y^2 \]

Since $s = \frac{2\pi^2}{{45}}g_{\ast s}T^3$, we can then write:

\[ {dY \over dx}  = -{k Y^2 \over x^{5/2}} \]

%%TYPO \[ {dY \over dx}  = -{\frac{2k}{3}Y^2 \over x^{5/2}} \]

where $k \equiv \frac{m_{\cal G}  2\pi^2\sigma_0 g_{\ast
s}m_{pl}}{1.67g_\ast^{1/2}45}$.

Here we will assume an average $g_\ast \approx g_{\ast s} \approx 50$.

Defining an order one parameter $\beta$ such that $\sigma_{0} = {{1}\over{(\beta TeV)^2}}$, $k \approx 4.5\cdot 10^{15} N \alpha / \beta^2 $.
%Here we will assume an average $g_\ast \sim g_{\ast s} \sim 50$

The solution to this equation is:

\[ Y_{final} = { 1 \over \frac{1}{Y_i}  + \frac{2k}{3}(\frac{1}{x_i^{3/2}} - \frac{1}{x_f^{3/2}})} \]

Notice a few properties of this solution. The present day
temperature is so low that $\frac{1}{x_f^{3/2}} \approx 0$. Hence
either the $\frac{1}{Y_i}$ term or the
$\frac{2k}{3}\frac{1}{x_i^{3/2}}$ term dominates, depending on $T_{RH}$. The $\frac{1}{Y_i}$ term dominates for $T_{RH} < .3 N \alpha / \beta^2$ MeV.  A reheat temperature in this
range would be inconsistent with nucleosynthesis, so we can ignore
this term.  Thus, $Y_f$  is determined by:

\[ Y_f = \frac{3x_i^{3/2}}{2k} \]

where $x_i = (m_{\cal G} /T_{RH})$.

In a general model where we do not fix the mass of the
${\cal G}$ baryon or the exact cross section, we can get an upper bound on $T_{RH}$ from our requirement that $Y_0 = {4.4\cdot 10^{-13} \over {N \alpha}}$:
\[T_{RH} > .008 \beta^{4/3} N \alpha \mbox{TeV}\footnote{When this number is larger than 1 TeV the calculation is not self consistent, because ${\cal G}$ baryon production is
thermal.}\] For reheat temperatures below this value, the ${\cal
G}$ baryons will dominate the universe. Thus we find a small
window $1 > {T_{RH}\over TeV} > .008  \beta^{4/3} N \alpha$, where non-thermal, symmetric ${\cal G} $ baryon production could account for the observed properties of dark
matter. In particular, for typical values $N \alpha \sim 10$ and $\beta \sim 1$, we find that this window has a width of about an order of magnitude. However, this range for the reheat temperature does not
conform to our prejudice that $m_{\cal G}$ is substantially larger
than $\Lambda_{\cal G}$.  We also note that there was no loss of
generality in our assumption that $m_X >> m_{\cal G}$.  If this
assumption is not valid, then $T_{RH}$ is quite low, and ${\cal
G}$ baryons would be overproduced as long as $m_X > m_{\cal G}$.

For $T_{RH} > 1$ TeV, the thermal relic abundance is too small to
account for the observed dark matter, but we can remedy this by
postulating an asymmetry.   The simplest possibility is that the
asymmetry is generated directly in the decay of the X particle, in
which case we have the standard result that

$$Y_0 = \epsilon_G {T_{RH}\over m_X} ,$$  where
\[ \epsilon_G \equiv \sum_\emph{f} B_\emph{f} \frac{\Gamma_X(X \longrightarrow \emph{f}) - \Gamma_X( X \longrightarrow \emph{\={f}})}{\Gamma_X} \]
 $\Gamma$ is a decay rate, $\emph{f}$ and $\emph{\={f}}$ are all
possible final states, and $B_\emph{f}$ is the total ${\cal G}$
baryon number of the final state $\emph{f}$. $\epsilon_B$ is the
corresponding asymmetry in ordinary baryon number.  In order to
match the observed dark matter density and the observed baryon
density, we need
$${\epsilon_G \over \epsilon_B} \approx {1\over 2 N\alpha} \times
10^{-2},$$ and $$ \epsilon_B {T_{RH}\over m_X}  \approx 8.6 \times
10^{-11}.$$

In our model $({{T_{RH}}\over m_X}\sim {\sqrt{m_X m_P}\over M})$\
\footnote{$M$ is the scale of irrelevant couplings of the $X$
particle to the standard model, which are responsible for the
decay.} is bounded from below by the requirements that the X
couplings to ordinary matter are at most Planck suppressed, and
that the $X$ is massive enough to produce the ${\cal G}$ baryon in
its decays.  Thus ${{T_{RH}}\over m_X} > ({N\alpha\over 2})^{1/2}
\times 10^{-3}$.

We see that the $\epsilon$ parameters must be very small in order
to account for the observed asymmetries.  In fact, small baryon
number violating branching ratios arise naturally if we assume
that X is a slow roll inflaton, $I$, with a ``natural" potential
of the form $\mu^4 f(I/m_P )$.  A theorem of Nanopoulos and
Weinberg\cite{nanowein} tells us that asymmetries can arise only
at second order in baryon violating couplings. Let us assume the
decay of the inflaton is mediated primarily via dimension $5$
operators.   Then even if the dimension five couplings involve CP
violation and baryon number violation, we will find that
$\epsilon_B \sim ({m_I \over m_P})^2 $.   We also have the order of
magnitude estimate ${T_{RH} \over m_I} \sim ({m_I \over
m_P})^{1/2}$, so that
$$Y_B \sim ({m_I \over
m_P})^{5/2}.$$ This will fit the observed baryon asymmetry if
$$m_I \sim 10^{-4} m_P$$  Note that this gives an inflation scale
$\mu$ close to the unification scale.

In this context we might attempt to explain the further
suppression ${\epsilon_G \over \epsilon_B} \sim 10^{-3}$ by
postulating that (perhaps as a consequence of the R symmetry
introduced in \cite{susycosmopheno}) the leading contribution to
the ${\cal G}$-baryon asymmetry comes from the interference of a
dimension $5$ and dimension $6$ coupling of the inflaton, and is
suppressed by a further power of $m_I \over m_P $.  This is off by
a factor of $10$ but our estimates are so crude that we can
consider this a success.

Indeed, we proposed this simple model not because we think it has
to be right, but to show that reasonable calculations of both the
dark matter and baryon abundances can be obtained for our new form
of dark matter.

To summarize, we probably need asymmetric production of ${\cal G}$
baryons to make them an acceptable dark matter candidate.   We
outlined a plausible model of asymmetric production in inflaton
decay, which could naturally explain both the baryon asymmetry of
the universe and the dark matter density.

\section{Conclusions}

We have shown that a baryon-like state of the new, strongly
interacting, ${\cal G}$ theory, which was introduced in
\cite{susycosmopheno} to implement Cosmological SUSY breaking, is
a promising dark matter candidate.  The new strong interaction
scale is around $1$ TeV and the ${\cal G}$ baryon mass is somewhat
higher, perhaps as high as the $15-18$ TeV needed to fit the Hess
data on photons from the center of the galaxy, if the explanation
for that data turns out to be dark matter annihilation. We saw
that this sort of baryon to interaction scale ratio was natural in
the context of large $N$ scaling with $N \sim 5-7$. While the
${\cal G}$ theory is probably not as simple as $SU(N)$ QCD, there
is reason to believe that reasonably large baryon masses are a
more general phenomenon.

This sort of dark matter candidate allows one to contemplate a
simple explanation of the dark matter to baryon ratio, since the
asymmetries in baryon and ${\cal G}$ baryon number might have the same
physical origin.  We need to explain a factor of order $10^3$ in
these asymmetries, in order to fit the data. We constructed a
plausible model in which both asymmetries are generated in
inflaton decay.   To explain the size of the asymmetries we
invoked the R symmetry of \cite{susycosmopheno} and dimensional
analysis.   There are more scenarios for baryogenesis in the
literature than there are authors on this paper, and it is
entirely plausible to us that a more elegant mechanism could be
found.  However, our simple model might work, and it might be the
right answer.

 Much more work needs to be done to sort out signatures of such
 hyper-strongly interacting dark matter, as well as to explore
 a variety of models for the production of baryon and ${\cal G}$ baryon
 asymmetries.   In addition, it will be necessary to find out more
 about the dynamics of the as yet mysterious ${\cal G}$ theory,
 which gives rise to these new particles.
\section{Acknowledgments}

We have benefitted from conversations with J. Primack, P.J. Fox,
M. Dine, and S. Dimopoulos.

The research of the authors was supported in part by DOE grant
number DE-FG03-92ER40689.

\end{document}